\documentclass[AMA,STIX1COL]{WileyNJD-v2}

\articletype{Research Article}%

\received{15 March 2023}
\revised{6 June 2023}
\accepted{6 June 2023}

\usepackage{amsmath,amsfonts}
\usepackage{array}
\usepackage[caption=false,font=normalsize,labelfont=sf,textfont=sf]{subfig}
\usepackage{verbatim}
\usepackage{graphicx}
\usepackage{booktabs}
\usepackage{multirow}
\usepackage{tabularx}
\usepackage{xcolor}
\usepackage{subcaption}
\begin{document}
\title{Distributed Edge Analytics in Edge-Fog-Cloud Continuum}

\author[1]{Satish Narayana Srirama $^{1}$}

\authormark{S. N. Srirama }

\address[1]{\orgdiv{School of Computer and Information Sciences}, \orgname{University of Hyderabad}, \orgaddress{\state{ Hyderabad 500 046}, \country{India}}}







\corres{Satish Narayana Srirama, \\ \email{satish.srirama@uohyd.ac.in}}


\abstract[Abstract]{

To address the increased latency, network load and compromised privacy issues associated with the Cloud-centric IoT applications, fog computing has emerged. Fog computing utilizes the proximal computational and storage devices, for sensor data analytics. The edge-fog-cloud continuum thus provides significant edge analytics capabilities for realizing interesting IoT applications. While edge analytics tasks are usually performed on a single node, distributed edge analytics proposes utilizing multiple nodes from the continuum, concurrently. This paper discusses and demonstrates distributed edge analytics from three different perspectives; serverless data pipelines (SDP), distributed computing and edge analytics, and federated learning, with our frameworks, MQTT based SDP, CANTO and FIDEL, respectively. The results produced in the paper, through different case studies, show the feasibility of performing distributed edge analytics following the three approaches, across the continuum.

}



\jnlcitation{\cname{%
\author{S. N. Srirama} 
} (\cyear{2024}), 
\ctitle{Distributed Edge Analytics in Edge-Fog-Cloud Continuum}, 
\cjournal{Internet Technology Letters}, \cvol{2024;00:1--6}.}

\maketitle


\section{Introduction}\label{sec:intro}

With increasing number of connected devices and ubiquitous Internet of Things (IoT) applications across different domains, the amount of data being generated is increasing in manifold dimensions and reaching the orders of zettabytes. This data collected from the sensors and IoT devices is traditionally moved to the cloud, where it is analysed, and the IoT equipment takes necessary actions based on the control signals resulting from this analysis. However, the increased latency, increased network load and compromised privacy issues associated with this Cloud-centric IoT (CIoT) model have led to the fog computing, where the data is analysed, closer to its source, by utilizing the proximal computational and storage capacity of devices such as network switches, gateways, routers etc~\cite{srirama2024decade}. Thus, with edge devices with sensing and actuation capabilities at the edge of the network/internet, fog nodes including the gateways, network devices, and private clouds in proximity, ultimately leading to the centralized public clouds, edge-fog-cloud continuum provides significant edge analytics capabilities for realizing interesting IoT applications. 

Data management in the edge-fog-cloud continuum includes data integration (e.g. federation, propagation, and consolidation), storage, preprocessing (e.g. anonymization, filtering, error detection, and encryption), batch and stream processing, and provisioning. Of these tasks, the processing of the data across the edge and fog devices is termed as edge analytics, which also includes preprocessing ~\cite{satyanarayanan2015edge}. Edge analytics provides preliminary insights from data streams, and it also protects cloud-based data storage against massive volumes, high data velocity and network congestion. Data management in edge-fog-cloud continuum and edge analytics have been studied extensively for over a decade~\cite{rosendo2022distributed}. From the literature, we can observe that most of the edge analytics tasks are performed on a single gateway/fog node. Even when there are more fog nodes involved, most of the tasks focused at applying different quality of service or quality of experience based scheduling strategies to find the ideal fog node to which the task is offloading~\cite{goudarzi2022scheduling}. Very few works, such as workflow-based applications, data pipelines, and applications requiring distributed data analytics, focused at taking advantage of multiple fog nodes
, thus collectively utilizing their resources for realizing the task~\cite{dautov2020stream, zhang2021edge, zhang2021survey}, mainly due to the nonexistence of easily adaptable frameworks. This paper calls these types of applications that take help of multiple fog nodes for edge analytics are based on \textit{distributed edge analytics}. The paper summarizes the state of the art in the data management and edge analytics domains and then summarizes our work related to distributed edge analytics. 

This paper takes its perspective on distributed edge analytics specifically in three directions. 1. Serverless data pipelines (SDP), where the data moves along the edge-fog-cloud continuum with interim edge analytics tasks being performed on the nodes across the continuum as serverless functions. 2. Distributed computing and data analytics performed on virtual clusters of resource constrained fog devices, collectively utilizing their resources, for realizing machine learning tasks. 3. Federated learning that iteratively trains using the local data samples at the fog devices, shares the trained model with centralised coordinator that aggregates and redistributes the model to the fog nodes for the next round. MQTT based SDP, CANTO and FIDEL frameworks are discussed that demonstrate the three distributed edge analytics perspectives, and the case studies along with their performance results produced in the paper show the feasibility of  the three approaches, across the edge-fog-cloud continuum.

The paper is organized as follows:
Section 2 discusses the state-of-the-art in the data management in the edge-fog-cloud continuum and edge analytics domains. Section 3 summarizes and demonstrates our serverless data pipelines work. Section 4 discusses our work on distributed machine learning on fog devices. Section 5 discusses and demonstrates our work on federated learning. Section 6 concludes the paper with future research directions. 

\section{State-of-the-art in data management in the edge-fog-cloud continuum and edge analytics}\label{sec:rel}

Data storage and delivery across edge-fog-cloud continuum is well studied and solutions such as Data as a Service (DaaS) have appeared. Badidi et al.~\cite{badidi2020fog} have discussed about DaaS as the fourth delivery model, in addition to IaaS, PaaS and SaaS, in fog-based environments, with data being delivered to the consumer on demand. Plebani et al.~\cite{plebani2018fog} proposed a DaaS-based solution to support data delivery in a fog computing environment, through formalization of data movement actions, context-based selection of valid movement actions and transformation and goal-based model at run-time for satisfying customers’ non-functional requirements.  

Regarding edge analytics, Dautov and Dietefano~\cite{dautov2020stream} proposed Apache Nifi data pipelines based distributed data processing on a cluster of edge devices. Poojara et al.~\cite{poojara2022serverless} studied three different approaches for deploying serverless data pipelines (SDP) across the edge-fog-cloud continuum, following the data handling mechanisms using Apache Nifi data pipelines, message queues, and object storage services such as AWS S3, respectively, while the data processing is based on serverless computing. 
This paper demonstrates SDP as an approach for distributed edge analytics in Section 3. 

Cloud, with its access to virtually infinite number of computational resources, is the ideal place for performing distributed data analytics. Fault-tolerant distributed data analytics framework such as Apache Hadoop MapReduce and its in-memory MapReduce alternative, Apache Spark, can be utilized for the IoT data analytics in the cloud. In addition, to deal with big streaming data of IoT, message queues such as Apache Kafka can be employed to buffer and feed the data into stream data processing systems such as Apache Storm and Apache Spark streaming~\cite{yang2017iot}. However, these approaches are applicable only for Cloud-centric IoT and are not suitable for the resource constrained fog computing setup.

Distributed edge analytics can also be performed on resource constrained fog devices, by collectively utilizing their storage and processing capability. However, there is no specific framework which can be used for generic distributed computing and data analytics on fog devices. Srirama et al.~\cite{srirama2021akka} adapted the Akka framework, based on actor programming model, for forecasting the sensor data using the ARIMA (Auto Regressive Integrated Moving Average) models, through distributed data analytics on Raspberry Pi node cluster. Tsai et al.~\cite{tsai2017distributed}  combined TensorFlow for splitting the application into smaller pieces, Docker for containerizing the pieces and Kubernetes for the cluster management, for performing distributed data analytics on fog devices. 

It is also common these days to utilize edge/fog devices for ML tasks. In most cases, the model is computed in the cloud, and the model is sent to the edge devices for drawing inferences. The public cloud providers have their own services such as AWS Greengrass and Azure IoT edge supporting these tasks. In addition to these, learning can also be performed on edge/fog devices. Generic distributable algorithms such as k-nearest neighbours (k-NN) and other special neural network methods, can directly be performed in resource constrained fog devices. There are also frameworks such as CANTO~\cite{srirama2023canto}, that can be used to train neural networks on fog nodes for performing distributed edge analytics. In addition to these, Federated Learning is emerging as another technology that utilizes the edge/fog resources for training the ML tasks, by only using the local data samples~\cite{zhang2021survey}. CANTO based distributed edge analytics training and federated learning are demonstrated in Sections 4 and 5.  Zhang et al.~\cite{zhang2021edge} provides a good review about performing distributed big data analytics and ML on edge devices.  

In summary, edge analytics and distributed edge analytics have been extensively explored in the literature. Rosendo et al.~\cite{rosendo2022distributed} provides a detailed literature review about edge analytics across the edge-cloud continuum. The current letter specifically focuses on distributed edge analytics, from the perspective of serverless data pipelines, distributed computing and edge analytics, and federated learning. Several works in the literature address each of these approaches separately, but no paper projects all the approaches together, putting the bigger scope of distributed edge analytics into perspective. The rest of the paper summarizes and discusses the key findings of our contributions in the distributed edge analytics domain. 

\section{Serverless data pipelines in edge-fog-cloud continuum } \label{sec:sdp}

Data pipeline concept allows data handling/moving from the source to the destination across multiple intermediate devices, each performing some functionality/process over the data in transmission. Data pipelines were already adapted in edge-fog-cloud continuum for edge analytics~\cite{dautov2020stream}. However, in IoT based applications, the edge analytics happening in the data pipelines is mostly event driven. For example, an action is performed, when the sensor data is over a certain range. The ideal framework in cloud computing domain that supports such event driven processing is the function-as-a-service or serverless computing. Thus, the serverless data pipelines (SDP) can be composed with data moving along the edge-fog-cloud continuum through data pipelines and interim data processing tasks performed on nodes as serverless functions. Here, we demonstrate the SDP designed with Message Queuing Telemetry Transport (MQTT). Further details about establishing the SDP with MQTT is provided in~\cite{poojara2022serverless}. 

\begin{figure}
\centering
\includegraphics[width=0.43\textwidth]{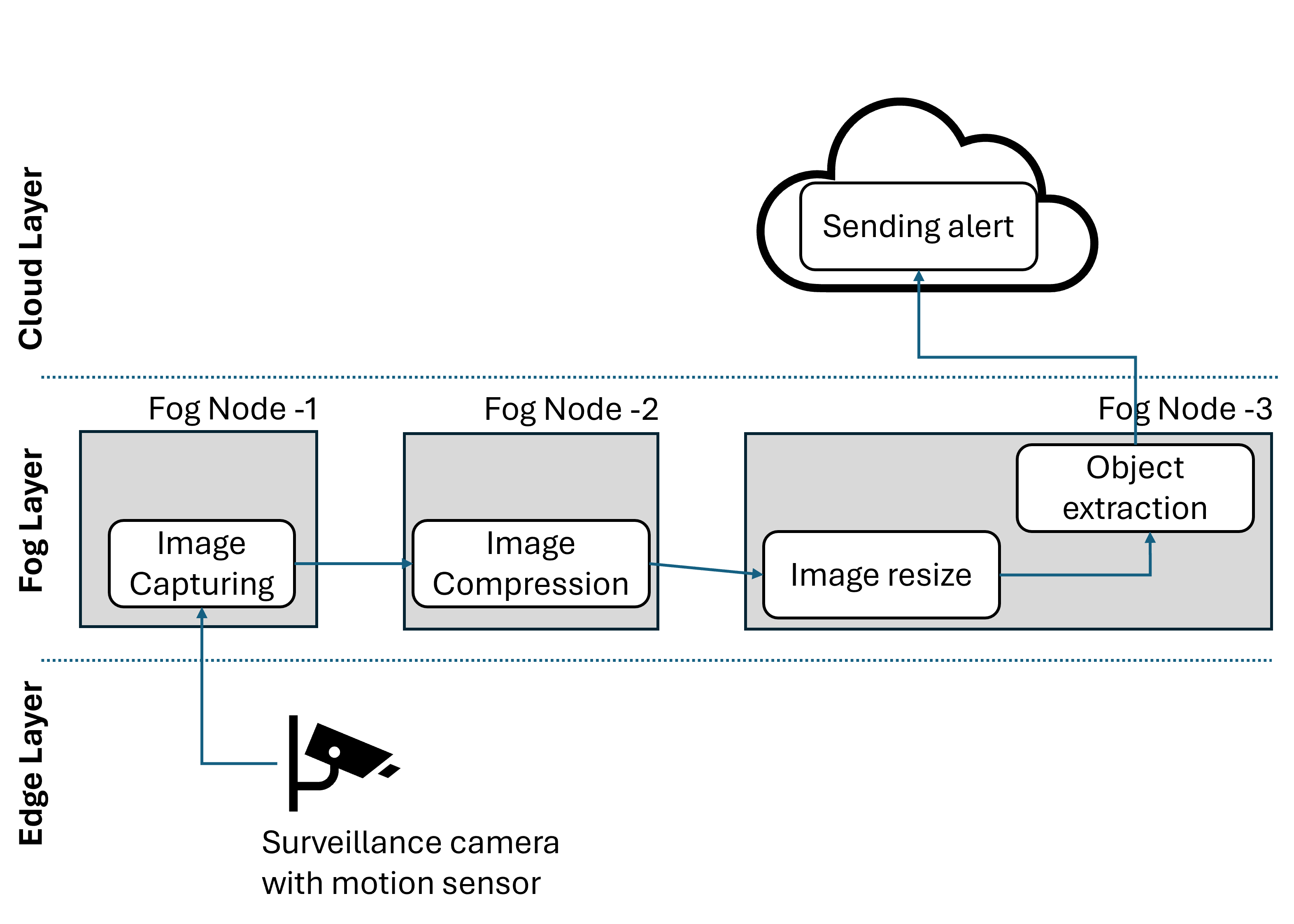}
\caption{Serverless data pipeline for IIoT surveillance}
\label{fig_SDPIIoT}
\end{figure}

For the demonstration, let us consider an Industrial IoT (IIoT) surveillance case study, where a surveillance camera’s photos are analysed for detecting human presence. The goal is to stop people accidentally coming close to certain equipment. The photos from this camera are captured based on motion sensor detection, these images are compressed, resized and then processed to extract the objects, as shown in Figure~\ref{fig_SDPIIoT}. If human presence is detected, the notice will be sent to the cloud for alert generation and storing the image for further analysis. Of these five processes, object extraction and alert sending are serverless functions. As shown in figure, in this SDP, four processes are performed on the fog nodes.

%

\begin{figure}[!t]
\centering
    \includegraphics[width=0.45\linewidth]{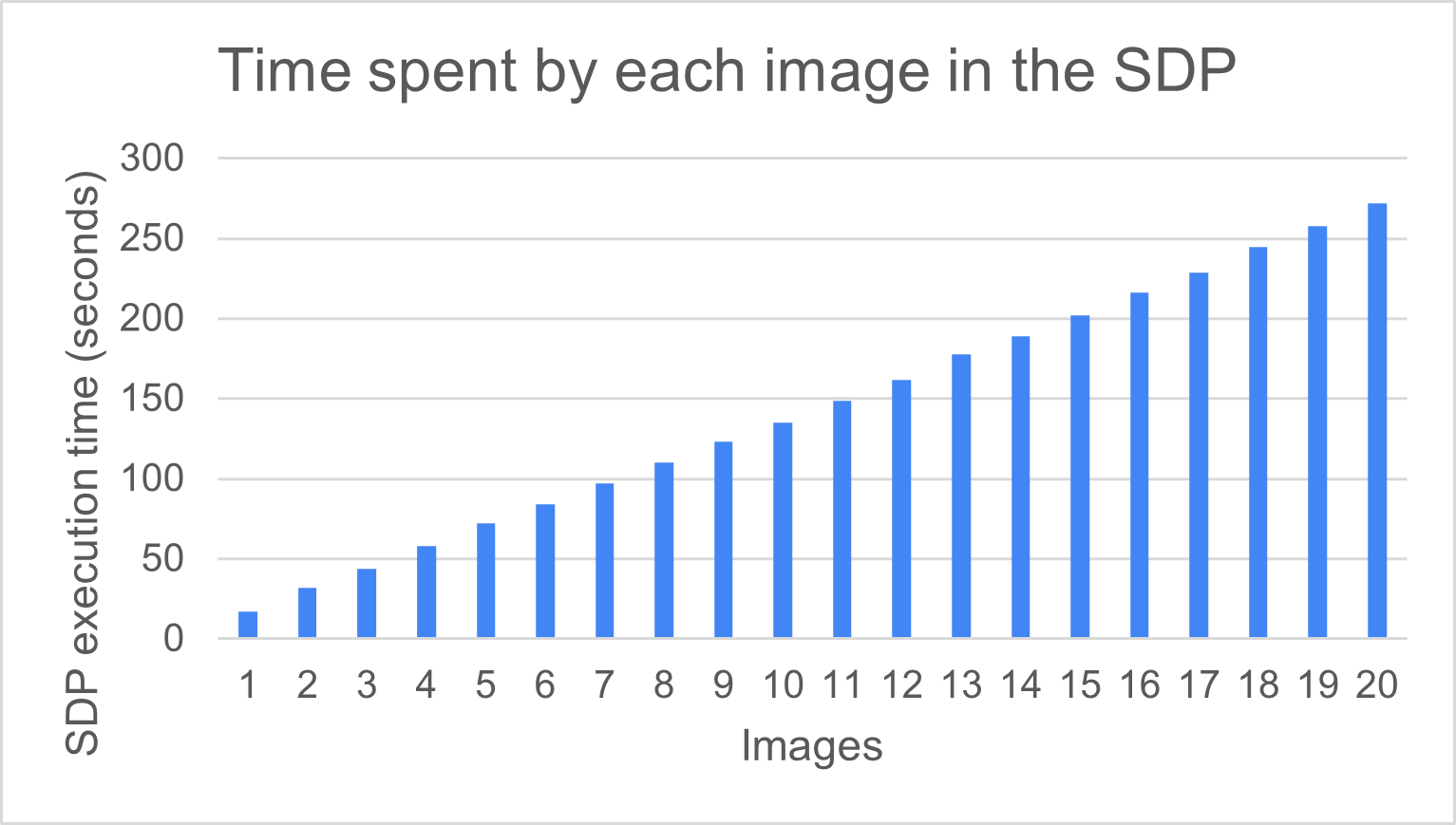} \hspace{1cm}
    \includegraphics[width=0.25\linewidth]{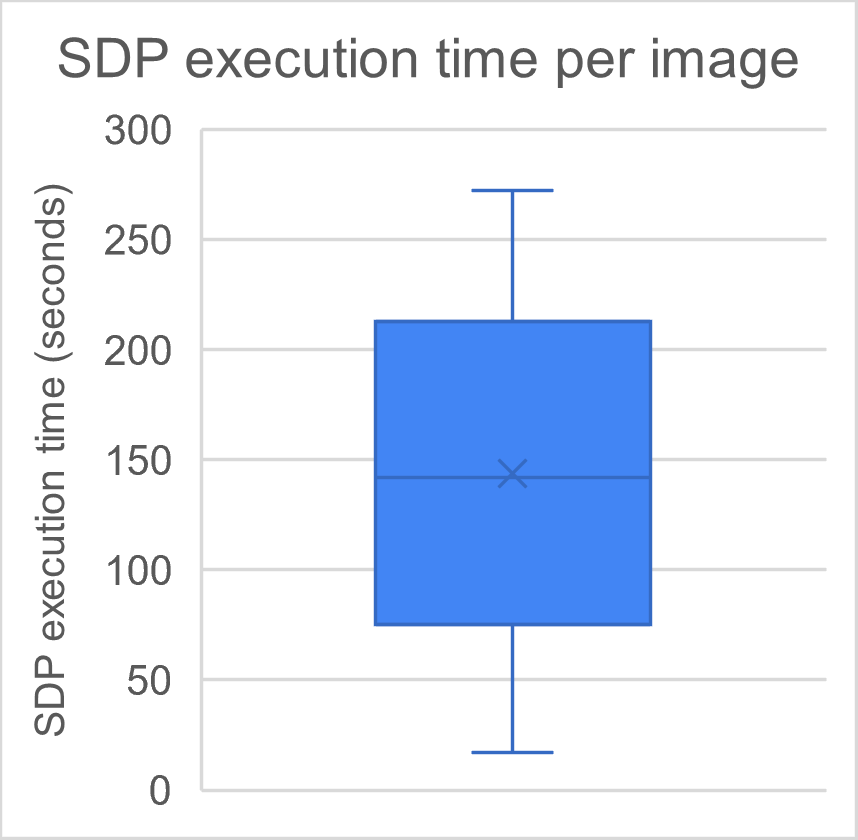}

\caption{(a) SDP completion time for each image (b) SDP execution time per image}
\label{fig:sdpInfo}
\end{figure}

We have realized the scenario on a fog layer with three Raspberry PI Model4 nodes, with Quad core Cortex-A72, 4 GB RAM. IR camera connected with the Fog Node -1, is used for capturing the images. 20 images were captured, with a 5 second gap between every capture, and are sent to the SDP. Figure~\ref{fig:sdpInfo} shows the SDP completion time for each image. Each image spent on average 143.6 seconds in finishing the complete pipeline, which may seem relatively high. However, we have to notice that the object extraction time for the image is about 14 seconds on the Raspberry PI and since the images are captured once every 5 seconds for this experiment, they are waiting in queue at the object extraction process. There is no failure in the experiment and all the images finished the SDP. In addition, from the SDP execution time of the first image, it takes 17 seconds for an image to complete the SDP, when all the queues are empty. Moreover, we also have to note that in real time, an image is captured only after a motion is detected. This analysis shows the feasibility of pipeline based distributed edge analytics task on edge-fog-cloud continuum. 

\section{Distributed computing and data analytics on the fog nodes} \label{sec:CANTO}

In the SDP, each of the pipeline process is performed on a single node. To support a task that is a bit more computationally heavy, we can make use of distributed processing across multiple fog nodes, by collectively utilizing their resources. Thus, for supporting parallel and fault-tolerant execution of the tasks, we adapted the Akka framework, and demonstrated performing wireless sensor data analytics on a cluster of fog nodes~\cite{srirama2021akka}. Akka framework is based on the Actor programming model, which was conceived as a universal paradigm for concurrent computation, also supporting resiliency and scalability. The Actor Model consists of a set of actors, which are isolated, concurrent, and interact solely through messages. Upon receiving a message, an actor performs a task or creates new actors. Actors distribute the workloads using a divide and conquer technique.

To support the further computational and communication intensive tasks such as machine learning, we extended our work and built CANTO as a general distributed computing framework that can train neural networks. The framework has provision to specify neural network parameters such as the neural network structure, dataset, the size of each dataset part, the activation function, learning rate, etc. More details about the framework and its usage can be obtained from~\cite{srirama2023canto}. 

The framework is demonstrated with the use case on IoT-based forest fire prediction. A three node Raspberry PI cluster is formed on which the CANTO framework is deployed. The forest fire prediction job is submitted from a laptop. The dataset~\cite{CortezForrestFire.24} consists of 10 features, that include meteorological parameters such as temperature, humidity, wind speed and rain along with other parameters. The classification with 4 classes denotes increasing levels of forest fire intensity ranging from 0 to 3. The dataset has 517 samples. After training the neural network with CANTO, we achieved an accuracy of 67\%. This is in line with the accuracy we achieved by running it on Google’s Colab, which is 65\%.

\section{Federated learning} \label{sec:FL}

\begin{figure}
\centering
\includegraphics[width=0.50\textwidth]{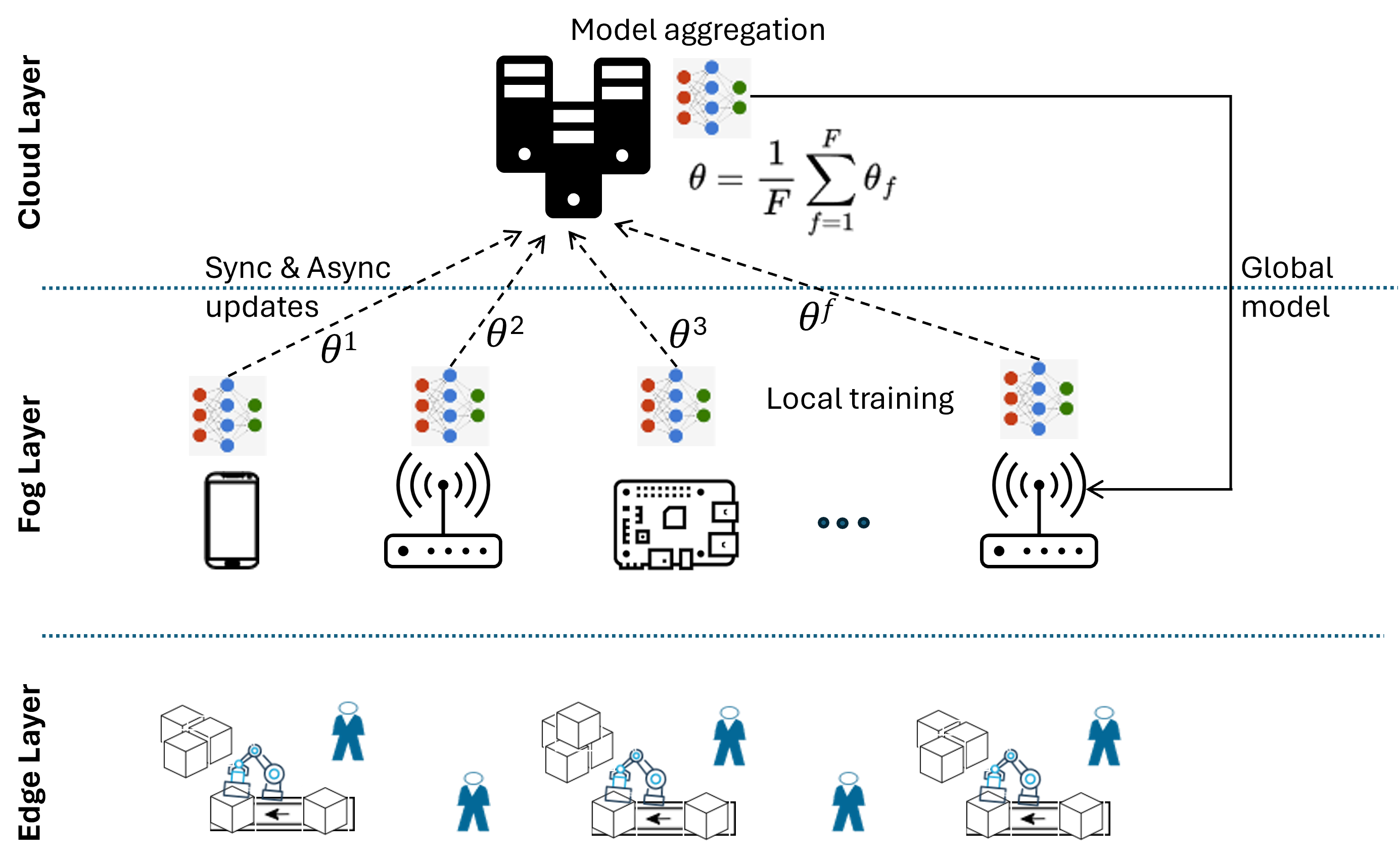}
\caption{Federated learning on the edge-fog-cloud continuum}
\label{fig_FL}
\end{figure}

Federated learning (FL) is a decentralized ML technique that can train a ML model on a group of fog devices, without sending raw data to a coordinating server on the fog or cloud layer. Each participating fog node generates/collects its own data on which the local model training is performed. The model along with its learned parameters is then shared with the server. Since the data is not directly shared, this approach preserves the data privacy. The sever node aggregates the individual models and generates the global model. The server coordinates with all participating clients and shares the global model for the next round of FL. The process repeats until the desired accuracy is achieved or end of training. Figure~\ref{fig_FL}, shows the FL in action. We have developed FIDEL, a federated learning framework, which supports both synchronous and asynchronous FL. In synchronous FL, the server waits until it receives the local models from all the participating fog nodes. In asynchronous FL, some of the fog nodes (called stragglers) can miss sending the local models (e.g. due to network/device failures) in some rounds and the server proceeds with global aggregation in regular intervals. Further details about FIDEL framework are at~\cite{kumar2024fidel}.

For the demonstration, let us consider another Industrial IoT (IIoT) case study, with Frequency-Modulated Continuous-Wave (FMCW) radars in a human–robot workspace. FMCW radars are effective at capturing a human position in the environment, which is critical for ensuring workers’ safety. By utilizing neural network models trained over the historic FMCW radars’ data to classify human safe distance, one can infer the current safety of the worker with extracted live radar data.

For the experiment, we have used 3 Raspberry PI nodes, and a laptop in the same network as the server node. We have used the dataset published at~\cite{8yqc-1j15-19}. The dataset contains 32,000 samples of FFT range measurements of 521 points in IIoT setup with 512 features and eight labels for human safety in terms of distance. The dataset is distributed across the nodes, randomly. 

%

\begin{figure}[!t]
\centering
    \includegraphics[width=0.40\columnwidth]{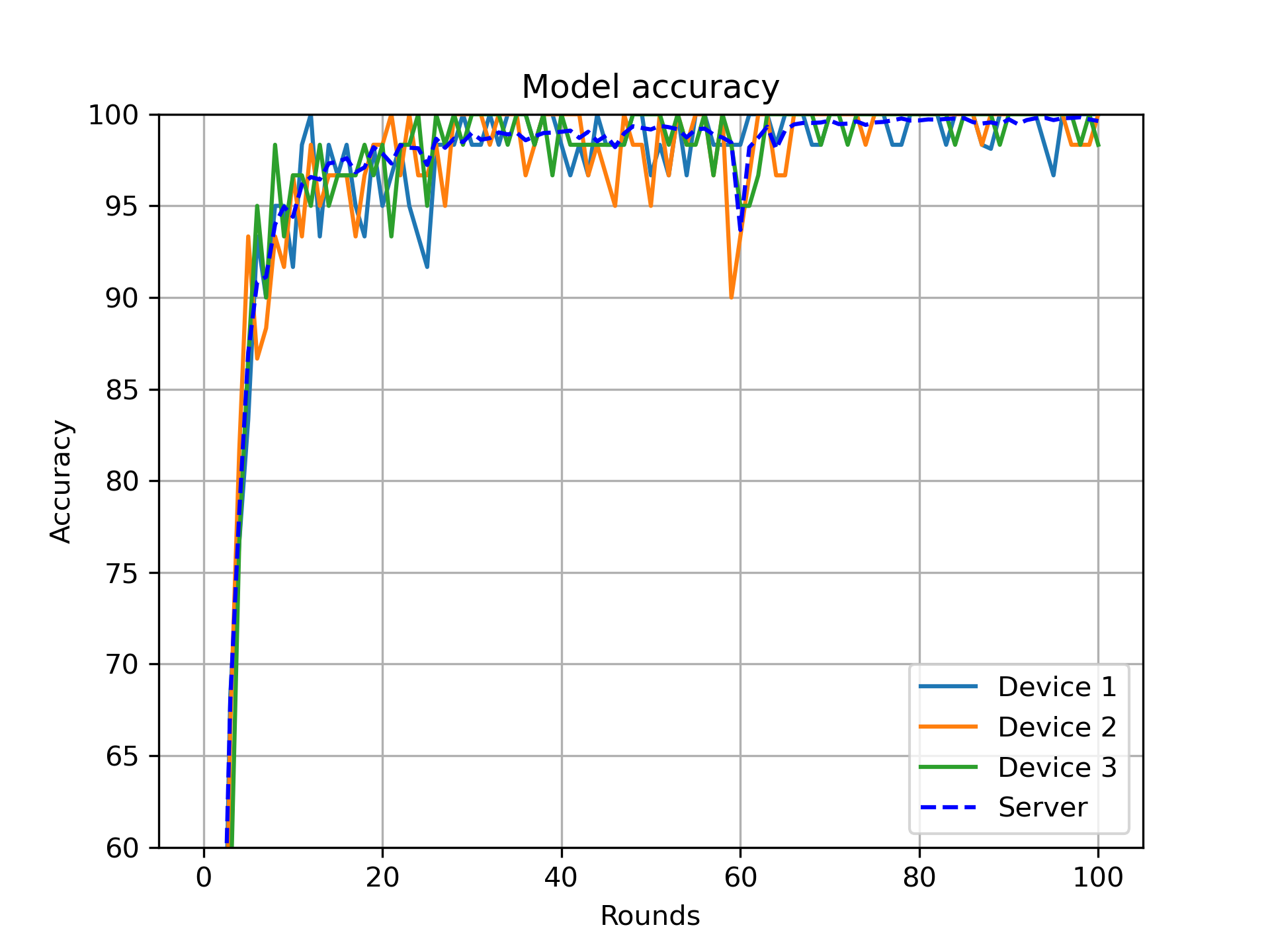}
\hspace*{0.25in}
    \includegraphics[width=0.40\columnwidth]{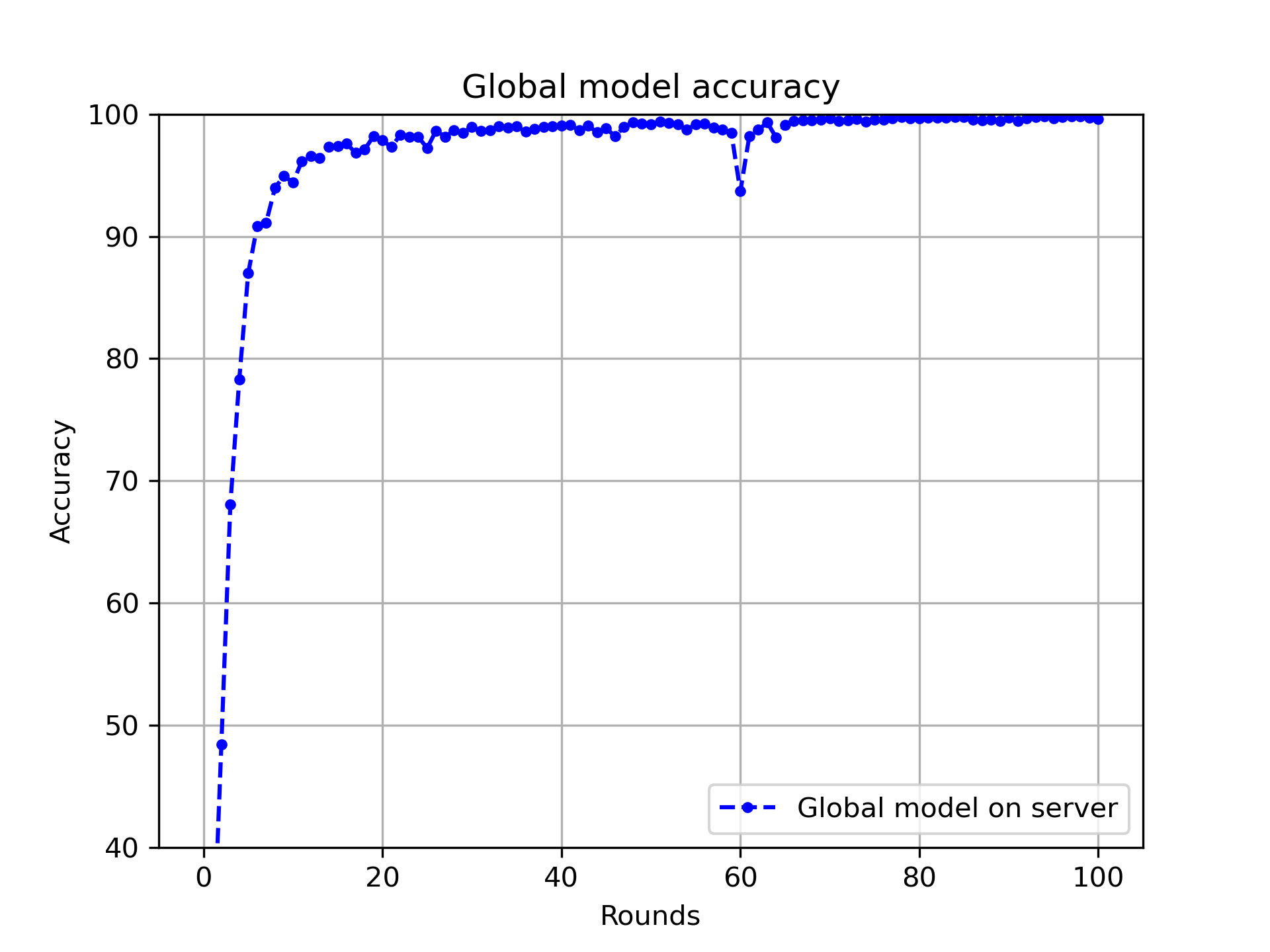}
\caption{FL accuracy over multiple rounds (a) Across all devices (b) On the test data with aggregated model on server}
\label{fig:FLInfo}
\end{figure}

The nodes perform federated learning by taking 60 samples per round for the local training, assuming data is generated every second and local training happens once per minute. The model considered is a neural network with one hidden layer with 32 neurons. The model has input layer with 512 neurons and output layer with 8 neurons. To use the complete dataset of 16000 training samples, we had 100 rounds of federated learning. Figure~\ref{fig:FLInfo} shows how the model accuracy increases as the federated learning rounds progress. This experiment shows, the distributed edge analytics is achieved through federated learning on edge-fog-cloud continuum.


\section{CONCLUSION}\label{sec:conclusion}
This paper discussed the state-of-the-art of data management, edge analytics and distributed edge analytics in the edge-fog-cloud continuum. The paper also promotes and demonstrated distributed edge analytics performed by taking three different perspectives: serverless data pipelines, distributed computing and edge analytics, and federated learning. SDP allows intermediate processing of data while it moves from edge to cloud. The second approach collectively utilizes the resources of multiple fog nodes through distributed computing, for handling relatively bigger edge analytics tasks. Federated learning preserves the privacy of data still performing machine learning tasks over it. Future research in this domain should focus at standardized and lightweight frameworks for distributed data analytics, their dynamic deployment, and providing incentives to the people/enterprises sharing their edge/fog resources.

\section*{Acknowledgement}
 This research is supported by SERB, India, through grant CRG/2021/003888. We also thank financial support to UoH-IoE by MHRD, India (F11/9/2019-U3(A)). The author also thanks his students, S. Poojara, S. Mirampalli, A. Kumar, D. Vemuri, S. Basak, A. Shekhar, who have contributed in producing the respective frameworks and results, over the years.

\bibliography{wileyNJD-AMA}%




\end{document}